\documentclass[aps,11pt,groupedaddress,superscriptaddress,nofootinbib]{revtex4-2}

\usepackage{graphics}%
\usepackage[english]{babel}
\usepackage[mathcal]{euscript}
\usepackage[latin1]{inputenc}
\usepackage{latexsym}
\usepackage{amsmath}    
\usepackage{graphicx}   
\usepackage{verbatim}  
\usepackage[dvipsnames]{xcolor}      
\usepackage{subfigure}  
\usepackage{bm}
\usepackage{graphicx}
\usepackage{amssymb}
\usepackage{bbm}
\usepackage{float}
\usepackage{slashed}
\usepackage{amsfonts}
\usepackage{euscript}
\usepackage{amsmath}    
\usepackage{mathrsfs} 
\usepackage{bbding}
\usepackage{pifont}
\usepackage{cancel}
\usepackage{tikz}
\usetikzlibrary{decorations.pathmorphing}
\usepackage{soul}
\usepackage[dvipsnames]{xcolor}
\usepackage[colorlinks=true,allcolors=NavyBlue,
linktocpage=true,pdfproducer=medialab,
pdfa=true]{hyperref}  

\begin{document}%
{\hspace*{13cm}YITP-21-46}
\title{\Large{Hearts of Darkness:\\ the inside out probing of black holes}}


\author{Ra\'ul Carballo-Rubio\footnote{\href{mailto:raul.carballorubio@ucf.edu}{raul.carballorubio@ucf.edu}}}

\affiliation{Florida Space Institute, University of Central Florida, 
 12354 Research Parkway, Partnership 1, 32826 Orlando, FL, USA}
\author{Francesco Di Filippo\footnote{\href{mailto:francesco.difilippo@yukawa.kyoto-u.ac.jp}{francesco.difilippo@yukawa.kyoto-u.ac.jp} }}
\affiliation{Center for Gravitational Physics, Yukawa Institute for Theoretical Physics, Kyoto University, Kyoto 606-8502, Japan}
\author{Stefano Liberati\footnote{\href{mailto:liberati@sissa.it}{liberati@sissa.it}}}
\affiliation{SISSA - International School for Advanced Studies, Via Bonomea 265, 34136 Trieste, Italy}
\affiliation{IFPU - Institute for Fundamental Physics of the Universe, Via Beirut 2, 34014 Trieste, Italy}
\affiliation{INFN Sezione di Trieste, Via Valerio 2, 34127 Trieste, Italy}

\begin{abstract}

\begin{center}
{Submitted 31 March 2021}
\end{center}
\centering\begin{minipage}{15cm}
Classical black holes shield us from the singularities that inevitably appear in general relativity. Being singularity regularization one of the main landmarks for a successful theory of quantum gravity, quantum black holes are not obliged to hide their inner core from the outside world. Notwithstanding the aforesaid, it is often implicitly assumed that quantum gravity effects must remain confined to black hole interiors. In this essay we argue in the opposite direction, discussing theoretical evidence for the existence of strong correlations between the physics inside and outside non-singular black holes. We conclude that astronomical tests of the surroundings of black holes can provide invaluable information about their so-far unexplored interiors.
\end{minipage}

\hspace*{-.52cm}
\begin{minipage}{17cm}
\begin{center}
   \vspace{1cm}
   \vspace{1.5ex}
    \hrule width \hsize  \kern .7mm \hrule width \hsize height 2pt 
       \vspace{1.5ex}
\textbf{\,Awarded an Honorable Mention in the 2021 Gravity Research Foundation Essays Competition 
}
   \vspace{1.5ex}
    \hrule width \hsize  \kern .7mm \hrule width \hsize height 2pt 
\end{center}
\end{minipage}

\end{abstract}

\maketitle
\newpage
\section{Introduction}

The growing observational evidence for black hole-like objects in accordance with general relativity predictions, also implies the existence of regions where the theory meets its demise.
A series of foundational results, starting from the 1965 Penrose's singularity theorem~\cite{Penrose1965}, shows that physically realistic initial conditions unavoidably produce a singularity as the endpoint of a gravitational collapse: a region where general relativity is not predictive and as such where spacetime is not defined.
Thus, at least close to a singularity, new physics appears to be required, and it is often implicitly assumed that any phenomena associated with it will be confined to a compact region inside the black hole.
The reason behind this implicit assumption is that there is no conceptual problem with general relativity that necessarily requires new physics around the horizon. However, it is worth noticing that long-range effects may well be a byproduct of introducing new physics, as the following analogy involving Newtonian gravity illustrates.

\emph{Lesson from Newtonian gravity:}
 Let us imagine to be confined in a region of the universe where it is not possible to directly observe high energy particles or strong gravity phenomena. Given that in this region Newtonian gravity represents an excellent approximation, we might naively accept the impossibility of testing general relativity even if we might have theoretically deduced it. 
 However, this does not need to be the case due to the existence of gravitational waves: produced in regions where strong gravity effects cannot be ignored, they can propagate to our ``Newtonian" region, so allowing us to probe the conjectured new physics. 
 Similarly, we cannot exclude {\em a priori}  the possibility that departures from general relativity produced at the core of black holes may generate long-range effects extending beyond their horizons. Of course, to properly explore such scenarios, we need some operative framework.
 
 To this end, in this essay we are going to construct a catalog containing all possible non-singular spacetimes with trapping, \textit{i.e.}~locally defined, horizons, and then use this catalog in conjunction with some additional considerations to determine whether long-range effects are to be expected. Our analysis can be schematically divided into three steps \cite{Classification,Pandora}:
\begin{enumerate}
\item[\textbf{1.}] \textbf{Classification:} The first part of our analysis consists in the geometrical characterization of the most generic non-singular spacetimes with trapping horizons. 

\item[\textbf{2.}] \textbf{Viability:} We then analyze the viability of each of these geometries according to a series of principles: plausible existence of a dynamical process leading to their formation, stability under perturbations, and compatibility with semiclassical physics.
\item[\textbf{3.}] \textbf{Phenomenology:} Viable geometries can be used to extract phenomenological implications of singularity regularization. 
\end{enumerate}

\section{Classification}
The central assumption we need to kickstart our analysis is that differential manifolds can provide a meaningful effective description of spacetime. While spacetime might in principle lose its smoothness in certain extreme situations, if that happens only in a bounded region, it always seems possible to devise a meaningful local effective geometry. 
Indeed, this expectation seems nowadays supported by several quantum gravity inspired models (see, \textit{e.g.} \cite{Kaminski:2010yz,Ashtekar:2011ni,Ashtekar:2015iza}).

The other assumptions we need to make are very straightforward. We assume that the spacetime is globally hyperbolic, so to guarantee a well defined causal structure,
and we also require the absence of singularities in two non-equivalent ways \cite{Geroch1967}: geodesic completeness and absence of curvature singularities.

\begin{figure}[!ht]
\begin{center}
\vbox{\includegraphics[width=0.8\textwidth]{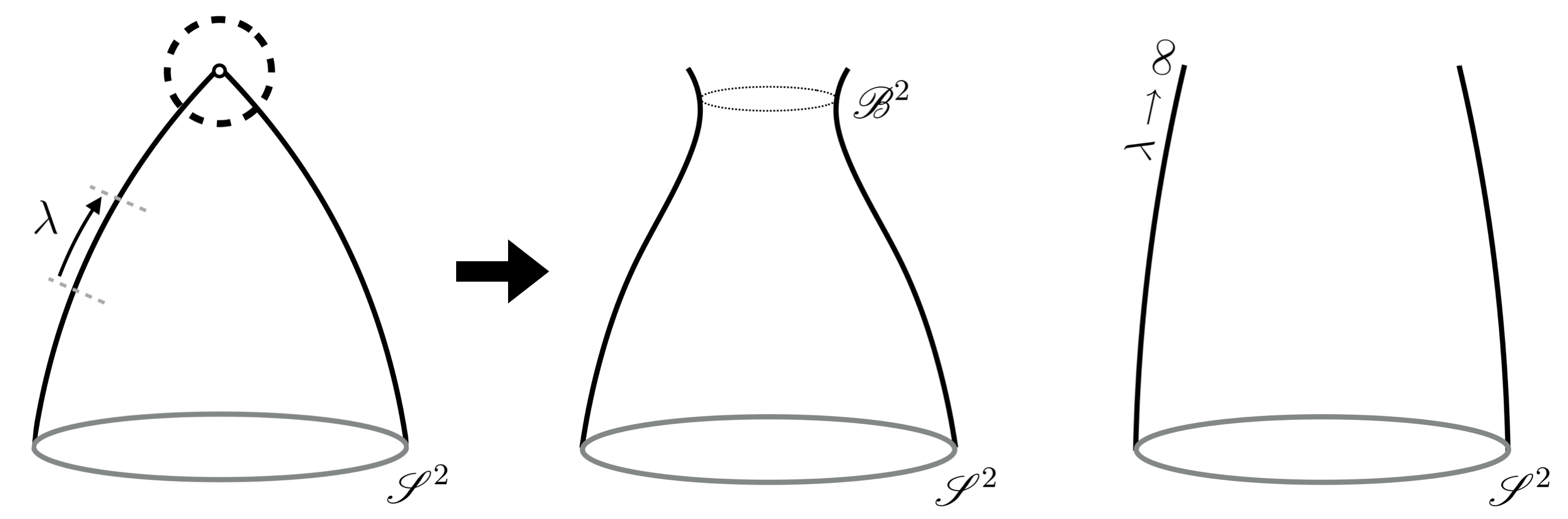}}
\caption{To guarantee geodesic completeness, either a defocusing point is created at a finite affine distance $\lambda$ (thus also creating the 2-surface $\mathscr{B}^2$) or at an infinite affine distance, or the focusing point is displaced to infinite affine distance. The figure on the right is compatible with both focusing and defocusing points at infinite affine distance.}
\label{fig1}
\end{center}
\end{figure}

Remarkably, these assumptions are sufficient to strongly constrain the most generic non-singular spherically symmetric black hole. The gist of the argument presented in~\cite{Classification,Pandora} by the current authors and another collaborator is the following. Outgoing spheres of light with origin in a trapped surface $\mathscr{S}^2$ \cite{Hawking1973}
are convergent. The Einstein field equations and the weak energy condition for matter fields lead to the development of a focusing point, which is incompatible with our assumption of geodesic completeness. Geodesic completeness can be recovered if the dynamics is modified in an open set around the focusing point, resulting in either a defocusing point at a finite or infinite affine distance or a focusing point displaced to an infinite affine distance (see Fig.~\ref{fig1}).

Combining these alternatives with the possible behavior of ingoing null geodesics lead to geometries that can be grouped in the following classes:

{\bf \emph{a.}} \emph{Evanescent horizons:} The singularity is replaced by an inner horizon and non-singular core. Inner and outer horizons merge in finite time due to a dynamical process that may or may not involve Hawking radiation. 

{\bf \emph{b.}} \emph{Hidden wormhole:} The singularity is replaced by a global or local minimum radius hypersurface, that is reminiscent of a wormhole throat hidden inside a trapping horizon.

{\bf \emph{c.}} \emph{Everlasting horizons:} Geometries in this class also have inner and outer horizons as those in the evanescent horizons class, although the two horizons never merge.

{\bf \emph{d.}} \emph{Asymptotic hidden wormhole:} This class can be obtained from the hidden wormhole class by pushing the wormhole throat to an infinite affine distance.

\section{Viability}
The geometrical analysis allows us to classify the most generic non-singular black hole spacetimes in only the aforementioned four classes of geometries. While this is already a very interesting result, we can constraint the possibilities even more by adding some physically motivated ingredients. 

{\bf \emph{a.}} \emph{Endpoint of gravitational collapse:} The lack of dynamical formation mechanisms can be used to discard some classes. The \textit{hidden wormhole} and \textit{asymptotic hidden wormhole} classes require topology change to be formed dynamically. However, topology change is not compatible with our assumption of global hyperbolicity~\cite{Geroch1967,Geroch1970,Borde1994}. Attempting to salvage global hyperbolicity by making the minimum radius hypersurface a local minimum rather than a global one turns out to be incompatible with the smoothness of the spacetime manifold.

{\bf \emph{b.}} \emph{Stability under perturbations:} Viable geometries must also be stable under perturbations. On purely geometrical grounds, it is possible to prove~\cite{Viability,Orimodel} that the inner horizon suffers from a linear instability under small perturbations in a short timescale (typically Planckian as measured by an asymptotic observer). While a complete non-linear analysis would require the specification of the dynamics of a particular quantum gravity theory, this result poses serious concerns regarding the stability of geometries with an inner horizon, namely those in the \textit{evanescent horizons} and the \textit{everlasting horizons} classes. Geometries in the \textit{evanescent horizons} class can sidestep this problem if the horizons merge in a timescale comparable with the instability timescale.    

{\bf \emph{c.}} \emph{Consistency with semiclassical physics:} Geometries in the \textit{everlasting horizons}, \textit{hidden wormholes} and \textit{asymptotic hidden wormholes} classes require that Hawking radiation switches off asymptotically. However, a vanishing Hawking temperature  
is incompatible with both the \textit{hidden wormholes} and the \textit{asymptotic hidden wormholes} classes, at least without the breakdown of standard quantum field theory.

\section{Phenomenology}
The features in common for the geometries in each of the four classes above allow us to draw some general conclusions about their phenomenological implications. The three classes that display self-consistency issues, namely the \textit{everlasting horizons}, \textit{hidden wormhole} and \textit{asymptotic hidden wormhole} classes, also share the property that geometries in these classes can be tuned\footnote{In general also these solutions can show long range effects, see \textit{e.g.}~\cite{Simpson_2019,Mazza:2021rgq}.} so to exactly match classical black hole spacetimes outside the trapping horizon, thus providing no phenomenological signatures. Similarly, long-living \textit{evanescent horizons}, albeit of problematic viability, may have no phenomenological signatures for a very long time, making them, in practice, indistinguishable from classical black holes. Interestingly, the only geometries that do not present any self-consistency issues {belong} to the short-living \textit{evanescent horizons} class. 
 
These geometries must display substantial deviations with respect to classical black holes, as trapping horizons must disappear in short timescales. However, the rapid disappearance of trapping horizons does not necessarily imply the sudden dispersion of the matter inside them. Actually, this would clearly contradict the observed lifespan of astronomical black holes. In contrast, the disappearance of the trapping horizon could instead cause a rearrangement of matter, which may for instance settle in the shape of an ultracompact object \cite{Barcelo:2007yk,Barcelo:2009tpa,Barcelo:2016hgb}. Searching for observable differences between ultracompact objects and classical black holes is then an even more motivated research area \cite{Phenomenology,Cardoso:2019rvt}.

\section{Conclusion}

We have discussed how minimal geometric assumptions suffice to construct an exhaustive catalogue of geometries describing in effective terms the regularization of black holes due to quantum gravity effects. 
This catalogue reveals a thought-provoking correlation between the physics inside and outside black holes: geometries describing non-singular black holes either display serious self-consistency issues in their interior, or substantial modifications to their exterior in timescales that make these modifications amenable to observational tests. This central conclusion of our analysis can be rephrased as the impossibility of modifying in an \textit{ad hoc} manner the interior of black holes without modifying their exterior, at least if certain basic physical principles are assumed. 

The reader may find this conclusion displeasing, or may object that it is a result of faulty assumptions, in particular the use of smooth geometries to describe the non-singular core of black holes. This possibility does not invalidate the main point in this essay, but further illustrates that keeping the physics outside black holes significantly untouched requires making specific assumptions about their interior, in particular that smooth spacetime ceases to exist there. Even if we cannot directly observe the interior of black holes, we should be able to follow the breadcrumbs that nature has left in the exterior for us.
\begin{acknowledgments}
\noindent
The authors would like to thank Matt Visser for several fruitful discussions.\\
FDF acknowledges financial support by Japan Society for the Promotion of Science Grants-in-Aid for Scientific Research No.~17H06359.\\
SL acknowledges funding from the Italian Ministry of Education and  Scientific Research (MIUR)  under the grant  PRIN MIUR 2017-MB8AEZ.\\

\end{acknowledgments}

\bibliographystyle{utphysics}
\bibliography{refs}
\end{document}